\documentclass{IEEEtran}

\usepackage{amssymb, amsmath}
\usepackage{color}
\usepackage{graphicx}
\usepackage{cite}
\usepackage{multirow}
\usepackage{threeparttable}
\usepackage{cases}
\usepackage{makecell}
\usepackage{adjustbox}
\usepackage{upgreek}
\usepackage{ragged2e} % for \RaggedRight macro
\usepackage{booktabs}
\makeatletter
\adddialect\l@ENGLISH\l@english
\makeatother
\usepackage[hidelinks]{hyperref}
\usepackage{tabularx,booktabs}
\graphicspath{{Figures/}}

\begin{document}

\title{\Large An Interface Method for Co-simulation of EMT Model and Shifted Frequency EMT Model Based on Estimation of Signal Parameters via Rotational Invariance Techniques} 
\author{Shilin~Gao,~\IEEEmembership{Member,~IEEE,}
   	Ying~Chen,~\IEEEmembership{Senior Member,~IEEE,}
   	Zhitong~Yu,
    Wensheng~Chen,
    Yankan~Song,~\IEEEmembership{Member,~IEEE}
    % Yuhong~Wang,~\IEEEmembership{Senior Member,~IEEE}
	\vspace{-1.9em}
\thanks{This work was supported by the Natural Science Foundation of China under Grant 52307127. \emph{(Corresponding author: Y. Chen)}}
\thanks{S. Gao and W. Chen are with the College of Electrical Engineering, Sichuan University, 610065, Chengdu, China (email: gaoshilin@scu.edu.cn, chenwensheng@stu.scu.edu.cn).} %, yuhongwang@scu.edu.cn
\thanks{Y. Chen, Z. Yu and Y. Song are with the Department of Electrical Engineering, Tsinghua University, 100084. (email: chen\_ying@tsinghua.edu.cn, yuzhitong@cloudpss.net, songyankan@cloudpss.net).}
}

\maketitle

\begin{abstract}
    The shifted frequency-based electromagnetic transient (SFEMT) simulation is much more efficient than traditional electromagnetic transient (EMT) simulation for ac grid. This letter proposes a novel interface for the co-simulation of the SFEMT model and the conventional EMT model. The fundamentals of SFEMT modeling are first derived. Then, an interface for the co-simulation of EMT and SFEMT models is proposed based on the estimation of signal parameters via rotational invariance techniques. Theoretical analyses and test results demonstrate the effectiveness of the proposed method.
\end{abstract}


\vspace{-1.5em}
\section{Introduction} \label{Sec1}

\PARstart{I}{n} traditional electromagnetic transient (EMT) simulation, a small time step is adopted to capture the dynamics with the minimum time-scale in the system, resulting in low computational efficiency. In fact, the dynamics of a large power system model involve multi-scale transients. Therefore, it is meaningful to develop a multi-scale simulation method that adopts different modeling methods and time steps for different subsystems. An idea of multi-rate EMT simulation with large step sizes for the ac grid and small step sizes for power electronics was proposed twenty years ago. However, for the simulation of ac grids, the time step-size is still small. 

\par To improve simulation efficiency, the shifted frequency-based EMT (SFEMT) simulation with a much larger step size is proposed \cite{ZhangP07, Strunz06}. Then, the co-simulation based on SFMET simulation and traditional EMT simulation is proposed in \cite{SHU110,tara14}, in which some subsystems of the power system are calculated by SFEMT simulation and the others are calculated by traditional EMT simulation. However, the interfaces for these co-simulation methods have a certain degree of accuracy loss. In these interfaces, the analytical signals for SFEMT simulation that are generated from real signals in the EMT simulation are not true analytical signals that are orthogonal to the original real signals.

\par Aiming at the problem of the existing interfaces, this paper proposes a novel interface for the SFEMT and EMT models. First, the general form of SFEMT modeling is derived, giving a guideline for the design of co-simulation interface. Then, an interface based on the estimation of signal parameters via rotational invariance techniques (ESPRIT) is proposed. The ESPRIT can accurately calculate the frequency, amplitude and phase of each component in an instantaneous signal with a short data window. According to the frequency, amplitude and phase of the original signal, the analytical signal that the SFEMT simulation needs can be constructed. The proposed interface is more accurate than the existing interfaces.

\par The letter is organized as follows. Section II derives SFEMT modeling fundamentals. The interface of co-simulation is designed and studied in Sections III and IV. Section V concludes.

\section{SFEMT Modeling Fundamentals} \label{Sec2} 

\subsection{General Form of SFEMT Modeling} \label{Sec2.1}
 
\par In EMT simulation, the dynamic equation of an electrical component can be generally expressed as:
 \begin{equation} \label{EqStateEquation}
    \frac{\mathrm d {x}(t)}{\mathrm d t}=Ax(t)+u(t)
 \end{equation}
 where $x(t)$ is the state variable, $A$ is the coefficient of $x(t)$, and $u(t)$ is the input. In \eqref{EqStateEquation}, the nonlinear part of the component dynamic is combined into $u(t)$, which is generally calculated by means of delay or prediction. For SFEMT modeling, a mathematical transformation $T[\cdot]$ (e.g., Hilbert transform) with the differential property ($T[\frac{\mathrm d x(t)}{\mathrm d t}] = \frac{\mathrm d T[x(t)]}{\mathrm d t}$) and linear property ($T[kx(t)] = kT[x(t)]$) is performed on both sides of \eqref{EqStateEquation} simultaneously to construct the adjoint system of \eqref{EqStateEquation}:
\begin{equation} \label{EqImaginary}
    \frac{\mathrm dT[x(t)]}{\mathrm d t}=A T[x(t)]+T[u(t)]
 \end{equation}

 \par Through \eqref{EqStateEquation}$+\mathrm j$\eqref{EqImaginary}, the system dynamic equation can be represented by analytical signal:
 \begin{equation} \label{EqStateEqLinear}
    \frac{\mathrm{d} \underline x(t)}{\mathrm{d} t}=A\underline x(t) + \underline u(t)
 \end{equation}
where $\underline x (t) $ and $ \underline u (t) $ are the analytical signals corresponding to $x (t) $ and $u (t) $, respectively. Then, by performing the frequency-shifting transformation on both sides of \eqref{EqStateEqLinear}, \eqref{EqComplexDifferential} is derived.
\begin{equation} \label{EqComplexDifferential}
    \frac{\mathrm{d} X (t)}{\mathrm{d} t}=AX(t)-\mathrm{j} \omega_{\mathrm{s}} X(t) + U(t)
\end{equation}
where $X (t) $ and $U (t) $ are the analytical envelopes corresponding to $x (t) $ and $u (t)$, respectively. By discretizing \eqref{EqComplexDifferential}, an SFEMT model for the component can be obtained.

\vspace{-0.5em}
\subsection{Impletation of Analytical Signal Consctruction}

\par To obtain a true analytical signal, the analytical signal $ \underline u (t) $ should not contain frequency components on the negative half axis of the spectrum \cite{Strunz06} for the use of a large time step-size. Otherwise, the signal component with negative frequency will become a signal component with a higher frequency after frequency shifting (e.g., $-65$ Hz to $-115$ Hz). Thus, to ensure that the signal $U(t)$ is a low-frequency envelope signal, the $T[\cdot]$ that is used to construct $\underline u(t)$ must also satisfy:
    \begin{equation} \label{SignalConstructionCondition}
        \mathcal{F}\left(T(u(t))\right)=-\operatorname{j} \operatorname{sgn}(f) \mathcal{F}(u(t)) 
     \end{equation}
where $\rm sgn(\cdot)$ is the sign function and $ \mathcal {F} (u (t)) $ is the spectrum of $u (t) $.

\par According to the above analyses, the guidelines of the analytical signal construction can be proposed. First, for an input $u(t)$ with a known analytical expression, the Hilbert transform can be used \cite{ZhangP07, Strunz06}. However, the convolution of the signal $u(t)$ and $h(t)$ in Hilbert transformation involves calculations from $-\infty$ to $\infty$, which cannot be implemented for signals with unknown analytical expressions. Then, for signals with unknown analytical expressions, the spectrum analysis-based methods (e.g., ESPRIT) can be considered, which satisfy \eqref{SignalConstructionCondition} and the two conditions in Section \ref{Sec2.1}.

\section{Interface Between EMT Model and SFEMT Model Based on ESPRIT} \label{Sec4}

\par In the muti-area co-simulation, the distributed transmission line model is widely used for the interface between different areas, which can also be used for co-simulation of the EMT and SFEMT models. The difference is that an interface between the real signal and the analytical envelope signal is needed in the co-simulation of the EMT and SFEMT models, which is detailedly elaborated as follows.

\subsection{Conversion Between Analytical Signal and Real Signal}

\par The real signal can be obtained from the analytical envelope signal by:
\begin{equation}
    x(t)=\operatorname{Re}\left(X(t) \mathrm{e}^{\mathrm{j} \omega_{\mathrm{s}} t}\right)
\end{equation}
where $x(t)$ is the real signal that the EMT model needs and $X(t)$ is the analytical envelope signal in the SFEMT model.

\par The analytical envelope signal that the SFEMT model needs can be calculated by:
\begin{equation}
    X(t) = \underline x(t) \mathrm{e}^{-\mathrm{j} \omega_{\rm s} t} =(x(t)+\mathrm{j} T[x(t)]) \mathrm{e}^{-\mathrm{j} \omega_{\rm s} t}
\end{equation}

\par Due to the inability to obtain the analytical expression of the voltage and current in EMT simulation, this paper proposes an analytic signal construction method based on the ESPRIT, which can accurately extract the spectrum of the real signal.

\subsection{Analytical Signal Construction Based on ESPRIT}

\subsubsection{Determination of the Number of Signal Components}

\par Assuming the sampling step size of the signal $x (t) $ is $\Delta t$, $x (t) $ can be discretized to $x(j\Delta t) = x_j$. For a data window with a length of $T_{\rm w}$, which contains $N=2n+1$ samples. The Hankel matrix $\boldsymbol X$ composed of signals within this data window can be represented as \cite{JainS105}:
\begin{equation}
    \boldsymbol{X}=
    \begin{bmatrix}
        x_{-{ n}} & x_{-{ n}+1} & \cdots & x_{0} \\
        x_{-{ n}+1} & x_{-{n}+2} & \cdots & x_{1} \\
        \vdots & \vdots & \ddots  & \vdots\\
        x_{0} & x_{1} & \cdots & x_{{ n}}
    \end{bmatrix}
\end{equation}
\par By performing singular value decomposition on the Hankel matrix, the singular value matrix  $\boldsymbol{\Sigma}$ can be obtained:
\begin{equation}
    \boldsymbol{\Sigma} = \operatorname{diag}(\sigma_1,~\sigma_2,\cdots,~\sigma_n)
\end{equation}
where every two consecutive singular values correspond to a signal component and the singular values corresponding to signal components are much greater than those corresponding to noises. According to this, the number of signal components $m$ can be determined.

\subsubsection{Calculation of the Frequency of Signal Component}

\par The Hankel matrix $\boldsymbol X$ can also be decomposed as \cite{Sheshyekani107}:
\begin{equation} \label{EqX_decomposition}
	\begin{aligned}
		\boldsymbol X & = \boldsymbol Z_{\mathrm L} \boldsymbol P \boldsymbol Z_{\mathrm R} \\
		& =
		\begin{bmatrix}
			1           & 1         & \cdots & 1 \\
			z_{1}       & z_{2}     & \cdots & z_{2m} \\
			\vdots      & \vdots    & \ddots & \vdots \\
			z_{1}^{n}   & z_{2}^{n} & \cdots & z_{2m}^{n} 
		\end{bmatrix}
		\begin{bmatrix}
			p_{1} & 0 & \cdots & 0 \\
			0 & p_{2} & \cdots & 0 \\
			\vdots & \vdots & \ddots & \vdots \\
			0 & 0 & \cdots & p_{2m}
		\end{bmatrix}
		\\
		& \quad  % 添加水平间距
		\begin{bmatrix}
			z_{1}^{-n}  & z_{1}^{-n+1}  & \cdots &  z_{1}^{0} \\
			z_{2}^{-n}  & z_{2}^{-n+1}  & \cdots &  z_{2}^{0} \\
			\vdots      & \vdots        & \ddots &  \vdots \\
			z_{2m}^{-n} & z_{2m}^{-n+1} & \cdots &  z_{2m}^{0} 
		\end{bmatrix}
	\end{aligned}      
	\end{equation}
where the $i$th element in $\boldsymbol Z_{\mathrm L}$ and $\boldsymbol Z_{\mathrm R}$ is $z_i = \mathrm{e}^{\mathrm j 2\uppi \frac{f_i}{f_{\rm sp}}}$, $z_i^{n} = \mathrm{e}^{\mathrm j 2\uppi n \frac{f_i}{f_{\rm sp}}}$, $f_{\rm sp}$ is the sampling frequency, $\boldsymbol P$ is a matrix composed of the frequency spectrum of each signal component (referred to as phasor). It is worth noting that $z_i$ can be solved based on the matrix pencil techniques \cite{Sheshyekani107}. Then, the frequency of each signal is calculated:
\begin{equation}
    {f}_{i}=\frac{\operatorname{Im}\left(\ln\left({z}_{i}\right)\right)}{2 \uppi \Delta t}
\end{equation}

\subsubsection{Calculation of Amplitude and Phase of Each Signal Component}

\par After obtaining $f_i$, $\boldsymbol Z_{\mathrm R}$ and $\boldsymbol Z_{\rm L}$ in \eqref{EqX_decomposition} can be calculated. By using the least squares method to solve \eqref{EqX_decomposition}, the phasor matrix $\boldsymbol P$ can also be obtained:
\begin{equation}
    \boldsymbol{P}=\operatorname{diag}\left(\left(\boldsymbol Z_{\mathrm{L}}^{\mathrm{H}} \boldsymbol Z_{\mathrm{L}}\right)^{-1} \boldsymbol Z_{\mathrm{L}}^{\mathrm{H}} \boldsymbol{X} \boldsymbol Z_{\mathrm{R}}^{\mathrm{H}}\left(\boldsymbol Z_{\mathrm{R}} \boldsymbol Z_{\mathrm{R}}^{\mathrm{H}}\right)^{-1}\right)
\end{equation}
where $\boldsymbol Z_{\mathrm{L}}^{\mathrm{H}}$ is the conjugate transposition of $\boldsymbol Z_{\mathrm{L}}$. The amplitude ${a}_{i}$ and phase ${\phi}_{i}$ of each signal component can be obtained by:
\begin{equation}
    \begin{aligned}
        {a}_{i}=2\left|{p}_{i}\right|, ~~{\phi}_{i}=\angle {p}_{i}
    \end{aligned}
\end{equation}

\subsubsection{Analytical Signal Consctruction}

\par When the frequency, amplitude, and phase of each signal component are obtained, the imaginary part signal $T\left( x(t)\right)$ in $\underline x (t) $ (denoted as $\hat x(t)$) can be accurately calculated:
\begin{equation}
    \hat x(t)=\sum_{i=1}^{m} a_{i}(t) \sin \left(2 \uppi f_{i} t+\phi_{i}(t)\right)
\end{equation}

\par Based on the original real signal and imaginary part signal, the analytical envelope signal $ X(t)$  is obtained:
\begin{equation}
    X(t)=\underline x(t)\mathrm{e}^{-\mathrm j \omega_{\rm s}t} = \left( x(t) + \mathrm j \hat x(t) \right) \mathrm{e}^{-\mathrm j \omega_{\rm s}t}
\end{equation}

\par It is worth noting that $\underline x(t)$ obtained by the ESPRIT is a true analytical signal, which only contains positive-frequency components. Then, the frequencies of the analytical envelope signal $X(t)$ obtained by frequency shift transformation are much lower than that of $x(t)$. On the contrary, the analytical signals obtained by methods of the delay transformation \cite{SHU110} and the second parallel simulation \cite{tara14} are virtual analytical signals. When there are harmonic components, if the methods in \cite{SHU110} and \cite{tara14} are used, the analytical envelope signal $X(t)$ in the interface will contain signal components with higher frequencies than the original signal.

\section{Case Studies} \label{Sec5}
\vspace{-0.25em}
\par
\begin{figure}
    \centering
    \includegraphics[width=2.75in]{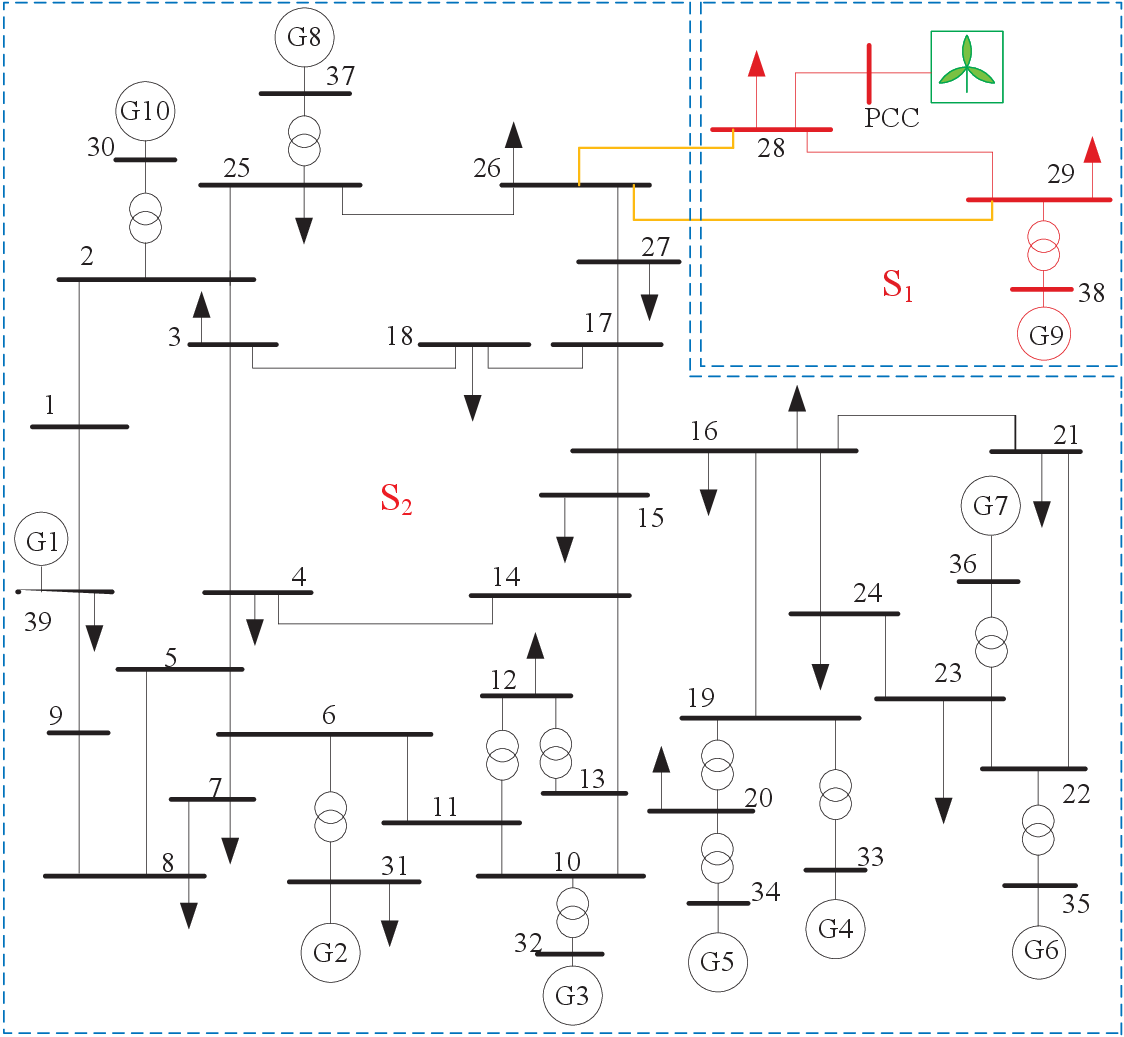}
    \vspace{-0.75em}
    \caption{Schematic diagram of the modified IEEE 39-bus power system.}
    \vspace{-1.5em}
    \label{Fig_IEEE39_WindFarm}
 \end{figure}

 The effectiveness of the proposed method is validated on a modified IEEE 39-node system with a wind farm, whose topology is shown in Fig. \ref{Fig_IEEE39_WindFarm}. The test system is divided into two subsystems ($\rm S_1$ and $\rm S_2$). The rated frequency of the ac grid is 50 Hz. According to the parameters of the interconnected transmission lines, the positive and zero traveling time of line 26-28 and 26-29 are calculated: $\tau_{2628}^{\mathrm p} = 612.13~\upmu$s, $\tau_{2628}^{0}=820.99~\upmu$s, $\tau_{2629}^{\mathrm p}= 807.23 ~\upmu$s and $\tau_{2629}^{0}=1.08$ ms.

 \par
\begin{figure}
    \centering
	\includegraphics[width=2.8in]{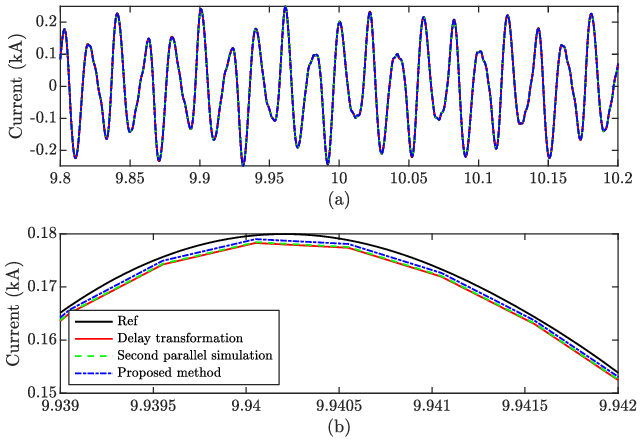}
    \vspace{-1em}
	\caption{Phase A current of the lines 26-25.}
	\label{Fig_IEEE39WindFarm_I2627}
    \vspace{-0.5em}
\end{figure}

 First, the test system is calculated by the co-simulation with the proposed interface, the delay transformation-based interface \cite{SHU110} and the second parallel simulation-based interface\cite{tara14}, respectively, where $\rm S_1$ and $\rm S_2$ are calculated by EMT simulation and SFEMT simulation, respectively. The time step-sizes for the two subsystems are 20 $\upmu$s and 500 $\upmu$s, respectively. The current of line 26-25 obtained by different methods is shown in Fig. \ref{Fig_IEEE39WindFarm_I2627}. Due to the integration of wind farm, a subsynchronous oscillation occurs and there are interharmonic components in the current. It can be found that the co-simulation based on the proposed interface is more accurate than the other two interfaces when there are harmonic components. The reason is that the imaginary part of the analytical signal obtained by the methods in \cite{SHU110} and \cite{tara14} does not satisfy the orthogonal relationship (i.e., \eqref{SignalConstructionCondition}) with the original real signal. The analytical signals obtained by these two methods still contain negative frequency components. After frequency shifting, the analytical envelope will contain signal components with higher frequencies. On the contrary, the analytical signal obtained by the proposed method only contains positive-frequency components. After frequency shifting, the frequency of the signal decreases, and the simulation accuracy will thus be improved. 

\begin{figure}
    \centering
	\includegraphics[width=2.6in]{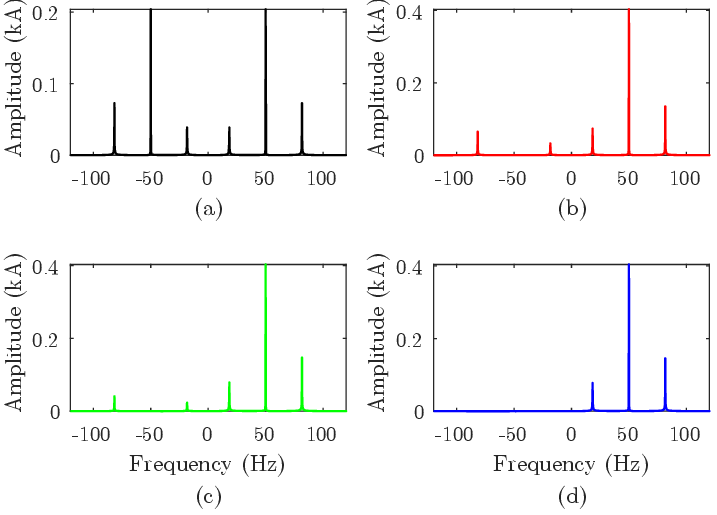}
    \vspace{-1.2em}
	\caption{Amplitude spectra of analytical signals obtained by different interfaces. (a) Original signal. (b) Delay transformation. (c) Second parallel simulation. (d) Proposed method.}
    \vspace{-1.5em}
	\label{Fig_Frequency_Spectrum_I2826}   
\end{figure}

\par Next, to further illustrate the advantages of the proposed method, the spectra of the analytical signals obtained from the three different interface methods are compared. For the current of line 26-28 that was obtained by different methods, the amplitude spectra of the analytical signals are shown in Fig. \Ref{Fig_Frequency_Spectrum_I2826}. It can be observed that the spectra of the analytical signals obtained with the interfaces of delay transformation \cite{SHU110} or the second parallel simulation \cite{tara14} contain negative frequency components. In contrast, the spectrum of the analytical signal obtained by the proposed method only contains positive frequency components. After frequency shifting, the obtained analytical envelope signal will become a low-frequency signal, and a larger simulation step size can be used.

\par Overall, when the voltage and current signals at the interface of the co-simulation of SFEMT models and EMT models contain harmonics, the accuracy of the ESPRIT-based interface is higher than that of the existing interfaces.

\section{Conclusion} \label{Sec6}

\par This letter proposes an interface for the co-simulation of EMT and the SFMET models. An analytical signal construction method based on ESPRIT is proposed for the interface, which improves the accuracy of co-simulation. Inspired by this letter, many other techniques in the signal process area may be introduced into the area of power system co-simulation. 

\bibliographystyle{IEEEtran}
\bibliography{References_v1.0}

\begin{thebibliography}{1}
\providecommand{\url}[1]{#1}
\csname url@samestyle\endcsname
\providecommand{\newblock}{\relax}
\providecommand{\bibinfo}[2]{#2}
\providecommand{\BIBentrySTDinterwordspacing}{\spaceskip=0pt\relax}
\providecommand{\BIBentryALTinterwordstretchfactor}{4}
\providecommand{\BIBentryALTinterwordspacing}{\spaceskip=\fontdimen2\font plus
\BIBentryALTinterwordstretchfactor\fontdimen3\font minus \fontdimen4\font\relax}
\providecommand{\BIBforeignlanguage}[2]{{%
\expandafter\ifx\csname l@#1\endcsname\relax
\typeout{** WARNING: IEEEtran.bst: No hyphenation pattern has been}%
\typeout{** loaded for the language `#1'. Using the pattern for}%
\typeout{** the default language instead.}%
\else
\language=\csname l@#1\endcsname
\fi
#2}}
\providecommand{\BIBdecl}{\relax}
\BIBdecl

\bibitem{ZhangP07}
P.~Zhang, J.~R. Marti, and H.~W. Dommel, ``Synchronous machine modeling based on shifted frequency analysis,'' \emph{IEEE Trans. Power Syst.}, vol.~22, no.~3, pp. 1139--1147, Aug. 2007.

\bibitem{Strunz06}
K.~Strunz, R.~Shintaku, and F.~Gao, ``Frequency-adaptive network modeling for integrative simulation of natural and envelope waveforms in power systems and circuits,'' \emph{IEEE Trans. Circuits Syst. I, Reg. Papers}, vol.~53, no.~12, pp. 2788--2803, Dec. 2006.

\bibitem{SHU110}
D.~Shu, X.~Xie, Z.~Yan, V.~Dinavahi, and K.~Strunz, ``A multi-domain co-simulation method for comprehensive shifted-frequency phasor {DC-grid} models and {EMT AC-grid} models,'' \emph{IEEE Trans. Power Electron.}, vol.~34, no.~11, pp. 10\,557--10\,574, Nov. 2019.

\bibitem{tara14}
J.~O. Tarazona, A.~T.~J. Marti, J.~R. Marti, and F.~A. Moreira, ``Shifted frequency {analysis-EMTP} multirate simulation of power systems,'' \emph{Electr Power Syst Res}, vol. 197, Aug. 2021.

\bibitem{JainS105}
S.~K. Jain, P.~Jain, and S.~N. Singh, ``A fast harmonic phasor measurement method for smart grid applications,'' \emph{IEEE Trans. Smart Grid}, vol.~8, no.~1, pp. 493--502, Jan. 2017.

\bibitem{Sheshyekani107}
K.~Sheshyekani, G.~Fallahi, M.~Hamzeh, and M.~Kheradmandi, ``A general noise-resilient technique based on the matrix pencil method for the assessment of harmonics and interharmonics in power systems,'' \emph{IEEE Trans. Power Del.}, vol.~32, no.~5, pp. 2179--2188, Oct. 2017.

\end{thebibliography}

\end{document}